\documentclass[aip, amsmath, amssymb, preprint, apl, longbibliography]{revtex4-1}

\usepackage[pdftex]{graphicx}
\usepackage{dcolumn}
\usepackage{bm}

\begin{document}

\title{Asymmetric Cryptography with Physical Unclonable Keys}

\author{Ravitej Uppu}
\email{r.uppu@utwente.nl}
\affiliation{Complex Photonic Systems (COPS), MESA+ Institute for Nanotechnology, University of Twente, P.O. Box 217, 7500 AE Enschede, The Netherlands.}

\author{Tom A. W. Wolterink}%
\altaffiliation[Current address: ]{Clarendon Laboratory, University of Oxford, Parks Road, Oxford OX1 3PU, United Kingdom.}
\affiliation{Complex Photonic Systems (COPS), MESA+ Institute for Nanotechnology, University of Twente, P.O. Box 217, 7500 AE Enschede, The Netherlands.}
\affiliation{Laser Physics and Nonlinear Optics (LPNO), MESA+ Institute for Nanotechnology, University of Twente, P.O. Box 217, 7500 AE Enschede, The Netherlands.}

\author{Sebastianus A. Goorden}
\altaffiliation[Current address: ]{ASML Netherlands B.V., De Run 6501, 5504 DR Veldhoven, The Netherlands.}
\affiliation{Complex Photonic Systems (COPS), MESA+ Institute for Nanotechnology, University of Twente, P.O. Box 217, 7500 AE Enschede, The Netherlands.}

\author{Bin Chen}
\affiliation{Department of Electrical Engineering, Eindhoven University of Technology, P.O. Box 513, 5600 MB Eindhoven, The Netherlands.}
\affiliation{School of Computing and Information, Hefei University of Technology, Hefei, China.}

\author{Boris \v{S}kori\'{c}}
\affiliation{Department of Mathematics and Computer Science, Eindhoven University of Technology, P.O. Box 513, 5600 MB Eindhoven, The Netherlands.}

\author{Allard P. Mosk}
\altaffiliation[Current address: ]{Nanophotonics, Debye Institute for Nanomaterials Research, Center for Extreme Matter and Emergent Phenomena, Utrecht University, P.O. Box 80000, 3508 TA Utrecht, The Netherlands.}
\affiliation{Complex Photonic Systems (COPS), MESA+ Institute for Nanotechnology, University of Twente, P.O. Box 217, 7500 AE Enschede, The Netherlands.}

\author{Pepijn W. H. Pinkse}
\email{p.w.h.pinkse@utwente.nl}
\affiliation{Complex Photonic Systems (COPS), MESA+ Institute for Nanotechnology, University of Twente, P.O. Box 217, 7500 AE Enschede, The Netherlands.}

\date{\today}

\begin{abstract}
{\bf Abstract:}
\linebreak
Secure communication is of paramount importance in modern society. Asymmetric cryptography methods such as the widely used RSA method allow secure exchange of information between parties who have not shared secret keys. However, the existing asymmetric cryptographic schemes rely on unproven mathematical assumptions for security. Further, the digital keys used in their implementation are susceptible to copying that might remain unnoticed. Here we introduce a secure communication method that overcomes these two limitations by employing Physical Unclonable Keys (PUKs). Using optical PUKs realized in opaque scattering materials and employing off-the-shelf equipment, we transmit messages in an error-corrected way. Information is transmitted as patterned wavefronts of few-photon wavepackets which can be successfully decrypted only with the receiver’s PUK. The security of PUK-Enabled Asymmetric Communication (PEAC) is not based on any stored secret but on the hardness of distinguishing between different few-photon wavefronts.
\end{abstract}

\maketitle

Secure communication has become of paramount importance in the internet era. 
The security is based on techniques that encrypt private messages from a sender (Alice) which can only be decrypted by the receiver (Bob) and not by any adversary (Eve). 
Symmetric cryptographic methods need an a priori exchange of secrets such as encryption keys and authentication keys between Alice and Bob.\cite{menezes1996} 
Asymmetric cryptography has been a major revolution in cryptography by overcoming the key distribution problem and allowing the encryption of messages to Bob with whom Alice does not yet share a secret. 
Asymmetric cryptography methods such as RSA and Diffie-Hellman key exchange use secret private keys (known only to its owner) together with public keys for security and thus overcome the necessity of a priori sharing a secret.\cite{diffie1976,rivest1978} 

The existing asymmetric cryptography methods face two issues. 
Firstly, they are not information-theoretically secure, i.e. the security relies on unproven mathematical assumptions such as the hardness of factorization or the computation of discrete logarithms. 
Secondly, digitally stored private keys are prone to stealthy copying (which is not detected by the key owner), thereby compromising the security. 
Over the last three decades, quantum physics has been exploited to create unconditionally secure cryptography methods such as Quantum Key Distribution,\cite{bennett1984,gisin2002,islam2017} Quantum Key Recycling,\cite{bennett2014,fehr2017,skoric2017} Quantum Secret Sharing,\cite{hillery1999} and Quantum Secure Direct Communication.\cite{long2002,deng2003} 
These methods utilize entanglement or the unclonability of unknown quantum states to avoid leakage of information to Eve or to detect Eve’s actions. 
Nevertheless, the practical implementation of these quantum methods again requires an authentication mechanism on the communication channel between Alice and Bob to prevent Eve from impersonating Bob. 
The standard approach for achieving this remains a priori sharing of a secret key, which to a large extent defies the purpose of a key exchange method. 
Indeed, public keys based on quantum states have been proposed as a way to fulfill all the security criteria.\cite{gottesman2001,nikolopoulos2008} 
However, the use of quantum states as public keys is highly impractical, since it requires long-term quantum storage, and has limited scalability in the number of keys.\cite{gottesman2001} 
Hence, there is still need for a practical asymmetric cryptographic method.

An important ingredient for a new cryptographic method are physical unclonable keys (PUKs), also known as physical unclonable functions (PUFs), which have been used as authentication tokens.\cite{pappu2002,buchanan2005,javidi2016,goorden2014} 
A PUK is a physical object with complex internal structure that is infeasible to copy due to the massive number of degrees of freedom that strongly affect its response to stimuli. 
PUKs that can be read out optically are readily realized in opaque scattering media (e.g. white paint, teeth and paper), which consists of vast numbers of randomly positioned particles. 
A recent development, Quantum Secure Authentication (QSA), verifies the authenticity of an optical PUK by querying it at the few-photon level.\cite{goorden2014,skoric2016} 
The security of QSA relies only on the hardness of cloning the PUK or its optical response. 

Here we combine PUKs and quantum cryptography: We introduce “PUK-Enabled Asymmetric Communication” (PEAC) that allows Alice to quantum-encrypt a message on few-photon wavefronts, which can be decrypted only with Bob’s PUK. 
The security of PEAC does not rely on mathematical assumptions or on the secure storage of secrets, but only on the hardness of distinguishing complicated high-dimensional quantum states. 
Importantly, PEAC requires only a one-way quantum communication channel between Alice and Bob. 

\begin{figure}
\centering
\includegraphics[width=\textwidth]{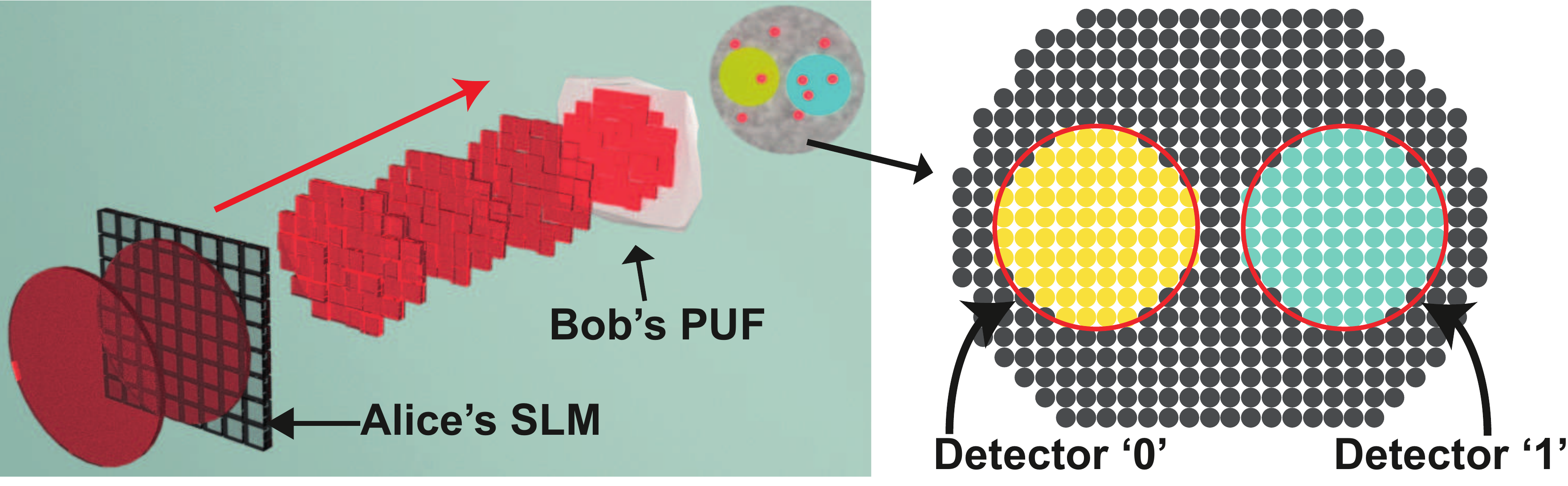}
\caption{{\bf PEAC Concept.} Using a SLM, Alice creates few-photon wavefronts that focus to one of the highlighted regions in the transmission of Bob’s PUK. The wavefronts are constructed as superposition of wavefronts known to focus to a spot in the detector area, indicated as gray dots. Photons in the highlighted yellow or blue spots are collected by a multimode fiber (core diameter indicated by the red circle) and detected on an avalanche photodiode D0 (‘0’) or D1 (‘1’), respectively. The received bit value is decided based on the difference in the number of photodetections in the two detectors.}
\end{figure}

Similar to many asymmetric cryptography methods, PEAC works with a public-private key pair. 
The private key is the infeasible-to-copy PUK held by Bob. 
The optical challenge-response characteristics of Bob’s PUK are utilized to form the digital public key. 
The public key is generated through a one-time optical characterization of the PUK as follows. 
A plane wave of light is programmed with the help of a digital Spatial Light Modulator (SLM) using complex wavefront shaping or digital phase conjugation in a setup illustrated in Fig. 1.\cite{mosk2012,rotter2017,huisman2014,wang2015} 
The SLM offers $K$ degrees of freedom in shaping the wavefront, i.e., $K$ independent phases can be programmed. 
The wavefronts are constructed such that the light transmitted through the PUK is focused to different locations, illustrated as the grid in the right panel of Fig. 1 (see Appedix A for details on experimental methods; Fig. 4, 5). 
Each distinct focus corresponds to a linearly independent incident field (wavefront). 
We separate  the  incoming wavefronts into two sets, each focusing to one of the highlighted regions in Fig. 1, and assign them as the bases $H_0$ and $H_1$ for transmitting classical bits ‘0’ and ‘1’, respectively. 
The two sets $H_0$ and $H_1$ together constitute the public key.

When any sender (for instance, Alice) wants to send a bit $b$ to Bob, they run the following procedure. 
Alice chooses as a random superposition $\psi$ of fields in $H_b (b \in \{0, 1\})$ and programs it in the SLM. 
A pulse of weak coherent light with low mean photon number $\langle n \rangle$ is patterned by the SLM and transmitted over the quantum channel to Bob. 
The quantum channel from Alice to Bob should be multimodal, with a capability to transmit $K$ spatial modes. 
The requirement of low $\langle n \rangle$ provides the security from eavesdropping thanks to quantum physical principles (discussed below). 
Bob uses two single-pixel detectors D0 and D1 that register the integrated photodetections over the areas corresponding to the spaces $H_0$ and $H_1$, respectively. 
The value of the received bit corresponds to the region with the maximum photodetections. 
The contrast between the signal in the detectors is high with a perfect concentration of light into the chosen area. 
However, practical limitations such as the noise in the source and the detectors and an incomplete control of the transmitted field result in a reduced contrast. 
Further, the partial transmission of the incident light by the PUK ($\approx 10\%$) leads to photon loss. 
The noise and losses lead to errors in the transmission of symbols from Alice to Bob, which necessitates error correction to make PEAC a functional communication scheme.

\begin{figure}
\centering
\includegraphics[width=\textwidth]{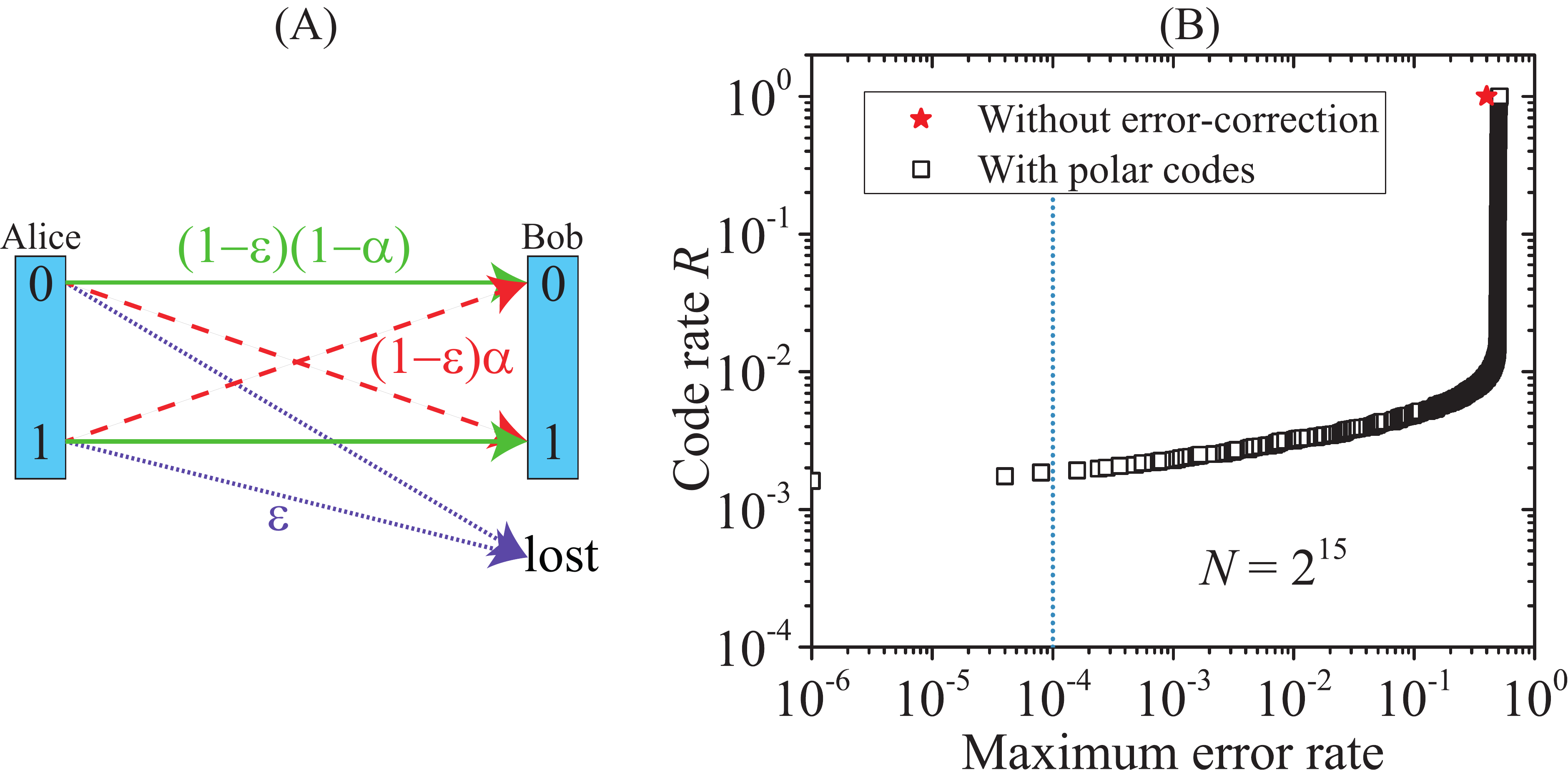}
\caption{{\bf Communication channel.} (A) Possible bit transmission outcomes: Correct transmission (solid green), loss (dotted purple) or flip of the bit (dashed red) are quantified by the two variables $\epsilon$ and $\alpha$,which correspond to bit loss and flip probabilities respectively. (B) the code rate $R$ as a function of the maximum A$\rightarrow$B error rate after application of the polar code (black squares right and left of the star, resp.). The channel error rate is estimated by transmitting $2^{15}$ bits, each carrying data i.e. code rate $R = 1$, over a channel with $\epsilon = 0.59$ and $\alpha = 0.43$ (red star). We design an error-correcting polar code for a codeword of length $N = 2^{15}$ transmitted over $N$ virtual channels. Simulations confirm the possibility to achieve bit error rate $< 10^{-4}$ for a code rate of $R \approx 0.002$, i.e. $M = R\cdot N \approx 2^6$ low-error decoded bits.}
\end{figure}

Our choice of error correction is guided by the level of channel noise. 
This can be quantified by the channel parameters shown in Fig. 2(A). 
The parameters $\alpha$ and $\epsilon$ characterize the bit-flip and bit-loss probabilities of the channel. 
The probability for a bit to be transmitted correctly is $(1-\epsilon) (1-\epsilon)$. 
At $\langle n \rangle = 33$ photons per wavefront with $K = 900$ degrees of freedom on the SLM, we find $\alpha = 0.43 \pm 0.01$ and $\epsilon = 0.59 \pm 0.01$. 
These channel parameters were estimated by transmitting $2^{15}$ pseudo-random bits from Alice to Bob. 
The error rate $\alpha$ is plotted as the red star in Fig. 2(B). 
To overcome the channel noise, we employ polar codes, which have been proven to be capacity-achieving.\cite{arikan2009} Conceptually, polar codes can reliably transmit $M$ bits of data by encoding them into a codeword of $N$ bits $(N > M)$. 
The other $N-M$ bits in the codeword are preassigned to a known value and encoded with $M$ data bits into an $N$ bit codeword. 
The operation of polar codes can be formulated through the construction of $N$ virtual channels with $M$ of them carrying data reliably. 
The ratio $R \equiv M/N$, called the code rate, quantifies the amount of information that can be transmitted. 
The achievable $R$ for a given channel is upper bounded by the channel capacity. 
Polar codes comprise an encoder-decoder system, which polarizes the $M$ data channels to have a vanishingly small error rate.
This improvement of the error rate in the data channels occurs at the cost of an increased error in the $N-M$ constructed noisy channels. 
Increasing the codeword length $N$ for a fixed $M$ improves the error rate of data channels but results in a lower communication speed, i.e. smaller $R$. 
Figure 2(B) shows the maximum error rate of the virtual channels in a codeword with $N = 2^{15}$ (shown as black squares) estimated using simulations based on density evolution methods.\cite{mori2009} 
A code rate $R = 2^{-9} \approx 0.002$ can be used with this codeword length to maintain a practical limit of bit error rate (BER) $< 10^{-4}$ (blue dashed line) for message transmission using polar coding (see Appendix C; Fig. 7). 

\begin{figure}
\centering
\includegraphics[width=\textwidth]{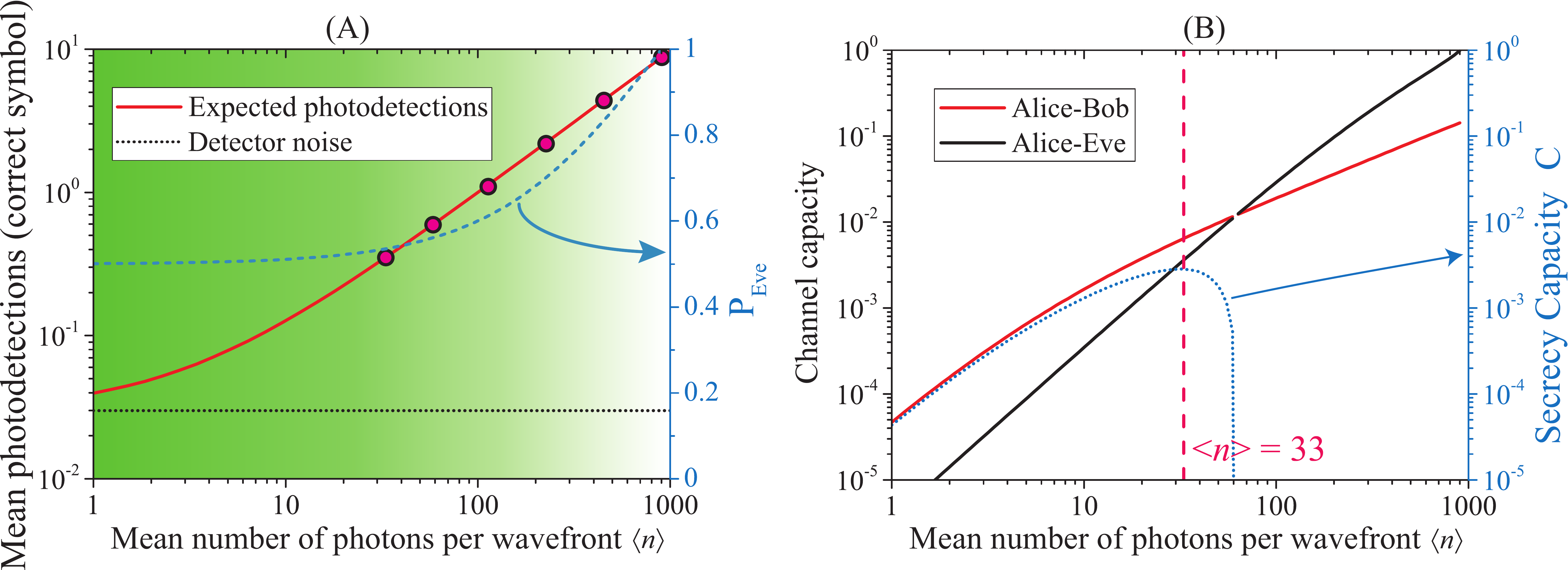}
\caption{{\bf Security.} (A) Measured number of photodetections per millisecond in the correct symbol with varying mean photon number $\langle n \rangle$ (magenta circles). The measured mean number of photodetections (error smaller than the circle size) agrees well with the estimated number of photodetections for an incident Poisson source with n photons (red curve). We operate well above the detector dark counts of 0.03 per millisecond (black dotted curve). Eve’s probability $P_{\textrm{Eve}}$ to guess the correct bit value is depicted as the blue dashed curve. Lower $P_{\textrm{Eve}}$ results in a higher level of security as highlighted with the darker shading. (B) depicts the channel capacity between Alice and Bob $I_{AB}$ (red curve) and between Alice and Eve $I_{AE}$ (black curve) as a function of $\langle n \rangle$. The secrecy capacity $C = I_{AB} - I_{AE}$ is also plotted (blue dotted curve) and reaches its maximum at $\langle n \rangle = 33$.}
\end{figure}

The security of PEAC relies only on one assumption: the quantum-physics-imposed difficulty of determining the used subspace $H_0$ or $H_1$ for $\psi$ at low photon number $\langle n \rangle$. 
Given that this assumption holds, the most generic attack by Eve is to estimate $\psi$ and infer the most likely subspace used in its construction. 
The most powerful state estimation is known to yield a fidelity $F=(\langle n \rangle+1)/(\langle n \rangle+K)$,\cite{bruss1999} from which we calculate the probability for Eve to correctly guess the subspace to be
\begin{equation}
	P_{\textrm{Eve}} \leq \frac{1}{q}+\frac{\langle n \rangle}{K} \frac{K-1}{K+\langle n \rangle}
\end{equation}
Here, $q$ is the number of subspaces; in our implementation $q = 2$. In the limit $K/\langle n \rangle \rightarrow \infty$ ,{\it i.e.} $\langle n \rangle \ll K, P_{\textrm{Eve}} \rightarrow 1/2$, which corresponds to a random guess. 
We define the security parameter $S \equiv K/\langle n \rangle$ to quantify the security, with $S > 1$ resulting in $P_{\textrm{Eve}} < 1$. 
Figure 3 shows the measured photodetections in the correct detector with varying $\langle n \rangle$. 
The baseline photodetections are the detector dark counts which impede Bob’s measurements at lower $\langle n \rangle$. 
At $\langle n \rangle = 33$, {\it i.e.} $S \approx 27$, the mean number of photodetections in the correct and wrong detector are 0.35 and 0.27, respectively. 
At $S \approx 27$, $P_{\textrm{Eve}} = 0.53$, very close to a random guess. 
The information shared between Alice-Bob and Alice-Eve can be quantified in terms of the channel capacities for Alice-Bob, $I_{AB}$, and Alice-Eve, $I_{AE}$, as shown in Fig. 3(B) (see Appendices for details). 
The secrecy capacity for the communication of information between Alice and Bob is $C = I_{AB} - I_{AE}$. 
When $C > 0$, the information shared between Alice and Bob is higher than that between Alice and Eve. 
In this scenario, standard privacy amplification techniques can be utilized to reduce Eve’s knowledge to practically zero. 
The error-correcting code has to be tuned such that the Alice$\rightarrow$Bob noise is corrected but not the Alice-Eve noise.
To ensure secure and error-free transmission of data between Alice and Bob, the designed polar codes should achieve BER $< 10^{-4}$ for the Alice$\rightarrow$Bob channel and an extremely noisy Alice$\rightarrow$Eve channel (ideally BER = 0.5). 
In our implementation with $C = 0.003$, we succeeded in designing a polar code with $R \approx 0.002$ that concurrently achieves an Alice$\rightarrow$Bob BER $< 10^{-4}$ and Alice$\rightarrow$Eve BER $> 0.12$ (see Appendix C; Fig. 8). 
After the error correction step, Eve’s partial information is reduced to zero using privacy amplification. 
Note that we assume the worst-case scenario in which Eve intercepts all the signal photons from Alice with perfect detectors. 
In contrast, the BER for the Alice$\rightarrow$Bob channel is estimated for the imperfect detection ($54\%$) and collection ($50\%$) efficiencies of the PUK transmission achieved in our setup. 

In conclusion, we demonstrate an asymmetric encryption scheme that does not require storage of any digital secrets. 
Its security is derived only from quantum physics and the technological infeasibility to make an optical device that copies the complex optical behavior of the physical key of the receiver. 
It can be deployed stand-alone or in tandem with classical cryptography; the latter case yields a strong multi-factor encryption with unprecedented security features. 
The proposed scheme can be readily implemented with available hardware, making it highly attractive for short-term realization. 
An important aspect in the physical realization of optical PUKs is the sensitivity of their optical response to environmental conditions. 
Optical PUKs that are robust against mechanical and thermal variations in the environment can be realized in ceramics, electrochemically-etched gallium phosphide and through laser micromachining of glass.\cite{schuurmans1999,zhang2016,*zhang2016err}
 PEAC can be extended to long distance communication by utilizing multimode fibers as transmission channels, which are currently employed for wavefront shaping and imaging.\cite{ploschner2015,amitonova2015}

\appendix
\section{Materials and Methods}
\begin{figure}[ht]
\centering
\includegraphics[width=\textwidth]{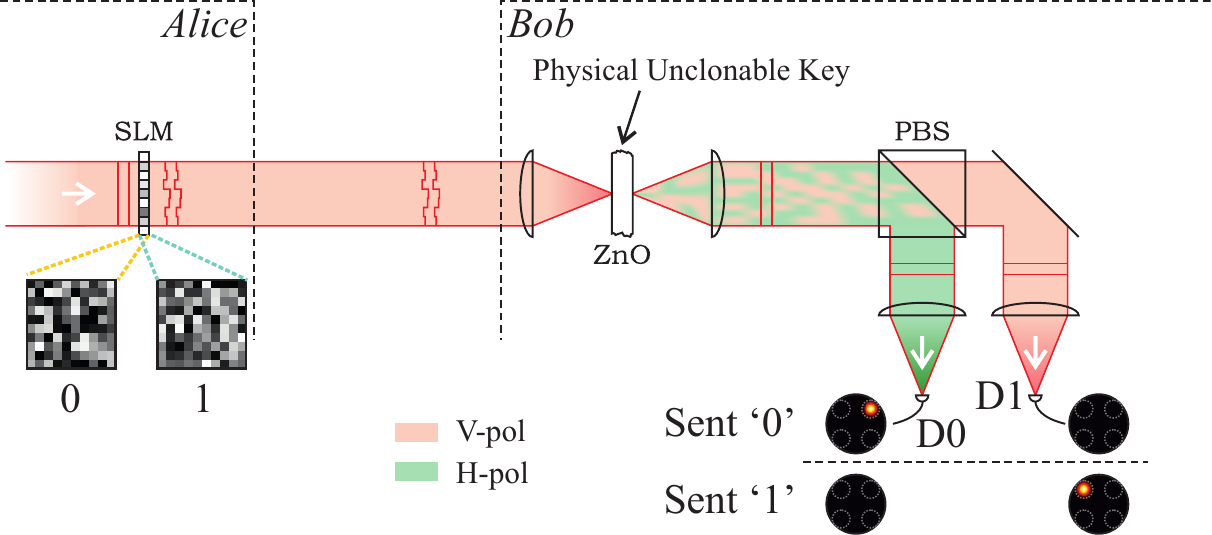}
\caption{{\bf Experimental setup.} The schematic of the experimental setup employed in PEAC is shown here. The incoming weak coherent light (vertically polarized) with an average photon number $\langle n \rangle$ is shaped using a liquid crystal spatial light modulator (SLM). The wavefront shaped light is transmitted to Bob located at 2 m distance. The received light is focused on Bob’s PUK composed of zinc oxide (ZnO) nanoparticles. The multiple-scattered light transmitted through the PUK is collected by a microscope objective. The horizontal and vertical polarization components in the transmitted light are separated using a polarizing beam splitter (PBS) and coupled into separate multimode fibers (200 $\mu$m core diameter) attached to photodetectors D0 and D1.}
\end{figure}

The experimental setup used in the implementation of PEAC is shown in Fig. 4. 
The wavefront of vertically polarized weak coherent light ($\lambda = 790$ nm; Coherent Mira) is shaped using a phase-only spatial light modulator (SLM; Hamamatsu X10468-02) divided into 900 segments. 
Each segment consists of a 10$\times$10 pixel group. 
The shaped wavefront is transmitted to Bob over a 2 m free-space channel. 
Bob uses a microscope objective (Zeiss 63x/0.95 NA) to focus the light onto the PUK. 
The PUK is a spray-painted slab (thickness 13 $\pm$ 1 $\mu$m) of zinc oxide (ZnO) nanoparticles on a 170 $\mu$m thick glass substrate.\cite{goorden2014} 
The light incident on the PUK is multiple-scattered as a result of the large thickness of the slab in comparison to the transport mean free path ($\ell_{\textrm{tr}} = 0.7 \pm 0.2 \mu$m). 
The multiple scattering scrambles the spatial modes and polarization of the incident light in transmission and reflection of the PUK, giving rise to a complex speckle pattern. 
The transmission speckle, comprising of $N_s$ spatial modes, is collected using a second microscope objective (Zeiss 64x/0.95 NA), which collects a fraction $f$ of the spatial modes. 
The three-dimensional multiple scattering of the incident light by the PUK results in an equal number of spatial modes in the horizontal and vertical polarization. 
We employ a polarizing beam splitter (PBS), to spatially separate the transmitted light into $fN_s/2$ horizontally-polarized spatial modes and $fN_s/2$ vertically-polarized spatial modes to achieve the spatial separation into orthogonal components (to be able to separate them easily without the detector fibers getting into each other’s way) required for PEAC as illustrated in Fig. 1. 
The bases $H_0$ and $H_1$ are chosen such that the transmitted light through the PUK focuses in either horizontal or vertical polarization, respectively. 
The cardinality of the union of basis sets should be close to the number of control channels, i.e. $|H_0 \cup H_1| \approx K$.
 
In the one-time optical characterization stage of the PUK, the spatially separated transmission of laser light ($\lambda = 790$ nm; power = 1 mW) is imaged onto two identical CCD cameras. 
The SLM is programmed using an iterative wavefront-shaping algorithm to focus to distinct locations on the CCD.\cite{mosk2012,rotter2017,huisman2014,wang2015,vellekoop2007,defienne2014,vellekoop2015}
The set of programmed SLM wavefronts for the distinct locations in the horizontal and vertical polarizations realize the bases $H_0$ and $H_1$, respectively.\cite{mosk2012,vellekoop2007} 
If the neighboring locations on the CCD are within the optical memory effect range of the PUK, the SLM patterns will have non-zero correlations.\cite{feng1988,freund1989,berkovits1994} 
We ensure minimal correlations between the basis patterns by selecting the largest subset of wavefronts with maximal Hamming distance ( $= K$, the number of SLM segments) to all basis wavefronts. 
After the characterization, the CCD cameras are replaced by lens-coupled multimode fibers (200 $\mu$m core diameter). 
The multimode fibers capture the whole speckle pattern and the integrated intensity or photodetections over the area is measured using photodiodes or avalanche photodiodes (D0 and D1) respectively. 
Figure 5(A) shows the measured photodiode reading with laser light when different patterns from bases $H_0$ and $H_1$ are incident on the PUK. 
A signal contrast, i.e. $[1 – (\textrm{correct detector signal})/(\textrm{wrong detector signal})]\cdot100\%$, of  $> 20\%$ (varies between $20\% - 53\%$ for different basis wavefronts) is observed across 350 independent patterns per basis. 
When random superpositions of the basis patterns are transmitted using weak coherent light, an average contrast of $30\%$ in the photodetections is observed as shown in Fig. 5(B). 

\begin{figure}
\centering
\includegraphics[width=\textwidth]{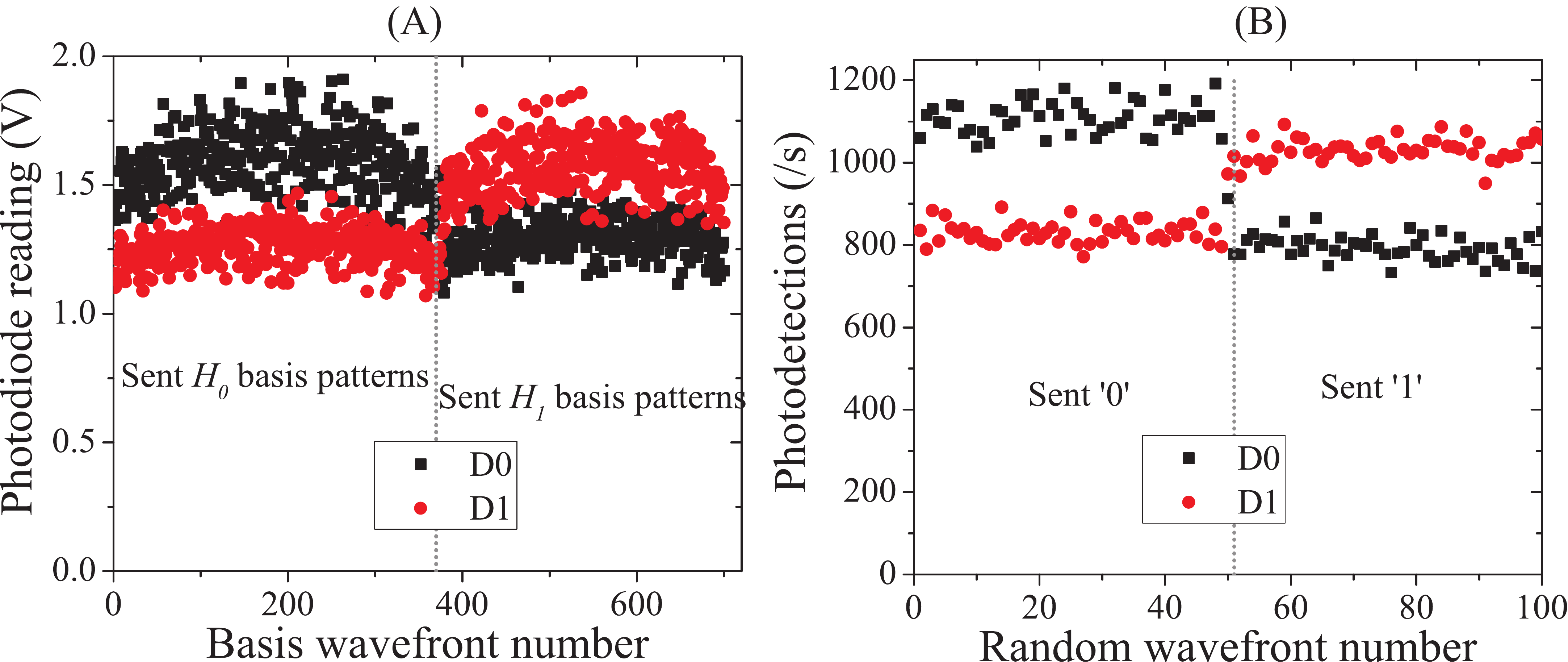}
\caption{{\bf Basis patterns.} (A) shows the one-time construction and calibration of the basis wavefronts of the PUK. The integrated intensities in detectors D0 and D1 are plotted for all the basis wavefronts of $H_0$ and $H_1$ patterned on the wavefront of coherent light using the SLM. (B) shows the contrast with weak coherent states with 100 different random superpositions of basis wavefronts, first 50 from $H_0$ and the rest 50 from $H_1$, are incident on the PUK.}
\end{figure}

The choice of spray-painted ZnO nanoparticles for PUK in our experiment arises from the ease in fabrication of multiply scattering material. 
However, the position of the nanoparticles in the ZnO slab is sensitive to small temperature variations, which becomes visible in the speckle pattern. 
The multiple scattering of light in the PUK is highly sensitive to the exact position of the nanoparticles.\cite{berkovits1994,berkovits1991} 
Multiple scattering media fabricated through electrochemical etching of gallium phosphide and laser micromachining of glass can be employed for PUKs that are robust against environmental fluctuations. 

\section{Security analysis}
We analyze the security of PEAC under the following assumption. 
It is infeasible to realize an optical device that projects any wavefront $\psi_b \in H_b$ to symbol $b$. 
Eve has knowledge of Bob's public key and the details of the symbol encoding scheme used by Alice and Bob. 
In this case, the most generic attack is to estimate the state $\psi$ and infer the most likely subspace used in its construction.
The most powerful state estimation method is known to yield a fidelity $F \leq (\langle n \rangle+1)/(\langle n \rangle+K)$, where $\langle n \rangle$ is the mean photon number per wavefront and $K$ is the number of degrees of control on the SLM.\cite{bruss1999} 
Let $\hat{\psi}$ denote Eve’s estimator for $\psi$. 
Then $F = |\langle \hat{\psi}| \psi \rangle |^2$. 
Let the correct subspace $H_x$, with $x \in \{0,1,\dots,q-1\}$, be spanned by an orthonormal system $\{\psi,v_2,v_3,…,v_{K/q}\}$. 
Averaged over random PUKs, the probability for Eve to correctly guess $x$ is given by
\begin{eqnarray*}
P_{\textrm{Eve}} &=& |\langle \hat{\psi}| \psi \rangle |^2 + \sum_{j=2}^{K/q}|\langle \hat{\psi} | v_j \rangle |^2\\
 &=& F+\sum_{j=2}^{K/q}|\langle \hat{\psi} | v_j \rangle |^2\\
 &\leq& F+\sum_{j=2}^{K/q} \frac{1}{K}\\
 &\leq& \frac{1}{q} + \frac{\langle n \rangle}{K} \frac{1-\frac{1}{K}}{1+ \frac{\langle n \rangle}{K}}.
\end{eqnarray*}
Here we have used $F \leq (\langle n \rangle+1)/(\langle n \rangle +K)$ and $|\langle \hat{\psi} | v_j \rangle |^2 \leq1/K$.
The latter inequality holds because an unconstrained average over the random vectors $|v_j \rangle$ in the $K$-dimensional space would yield $1/K$, whereas our average excludes the favorable direction $\psi$. 
The binary case realized in our experiments can be retrieved by setting $q = 2$. For $\langle n \rangle \ll K$, $P_{\textrm{Eve}} \rightarrow 1/q$, which is equivalent to a random guess of the correct basis.

Using the above result, we can calculate the channel capacity between Alice and Eve $I_{AE}$ to be
\begin{equation*} 
I_{AE}=1-h(P_{\textrm{Eve}} ),
\end{equation*}
where the binary entropy function is $h(x)=x \log_2  1⁄x+(1-x) \log_2  1⁄((1-x))$.  
The channel capacity for a binary symmetric erasure channel between Alice and Bob $I_{AB}$ can be written using the channel parameters $\epsilon$ and $\alpha$ as
\begin{equation*}
I_{AB}=(1-\epsilon)(1-h(\alpha)).
\end{equation*}

\section{Polar codes}
Polar codes are linear block-error-correction codes, which are mathematically proven to be capacity achieving for memoryless binary channels.\cite{arikan2009} 
Let’s consider a binary channel $W$ which maps input $X \in \{0, 1\}$ to output $Y$ with channel capacity $I(W) = I(X;Y)$, see Fig. 6(A). 
Conceptually, given $N$ copies of the binary channel $W$, polar codes synthesize $N$ virtual channels $W_N^{(i)}  (i \in \{1,2,\dots,N\})$, such that the channels with indices from a set $F$ are “bad” (completely noisy with $I(W_N^{(i)})\rightarrow0,\forall i\in F)$ and the channels with indices from the complement of set $F$ are “good” (nearly noiseless with $I(W_N^{(i)} ) \rightarrow 1,\forall i\in\{1,2,\dots,N\} \backslash F)$ i.e. the virtual channels are polarized. 
In the limit of $N \rightarrow \infty$, the fraction of good channels approaches the channel capacity $I(W)$, i.e. the Shannon bound. 
Information is transmitted over the good channels, while fixed values (frozen bits) known both at encoder and decoder are assigned to the bad channels. 

\begin{figure}
\centering
\includegraphics[width=0.7\textwidth]{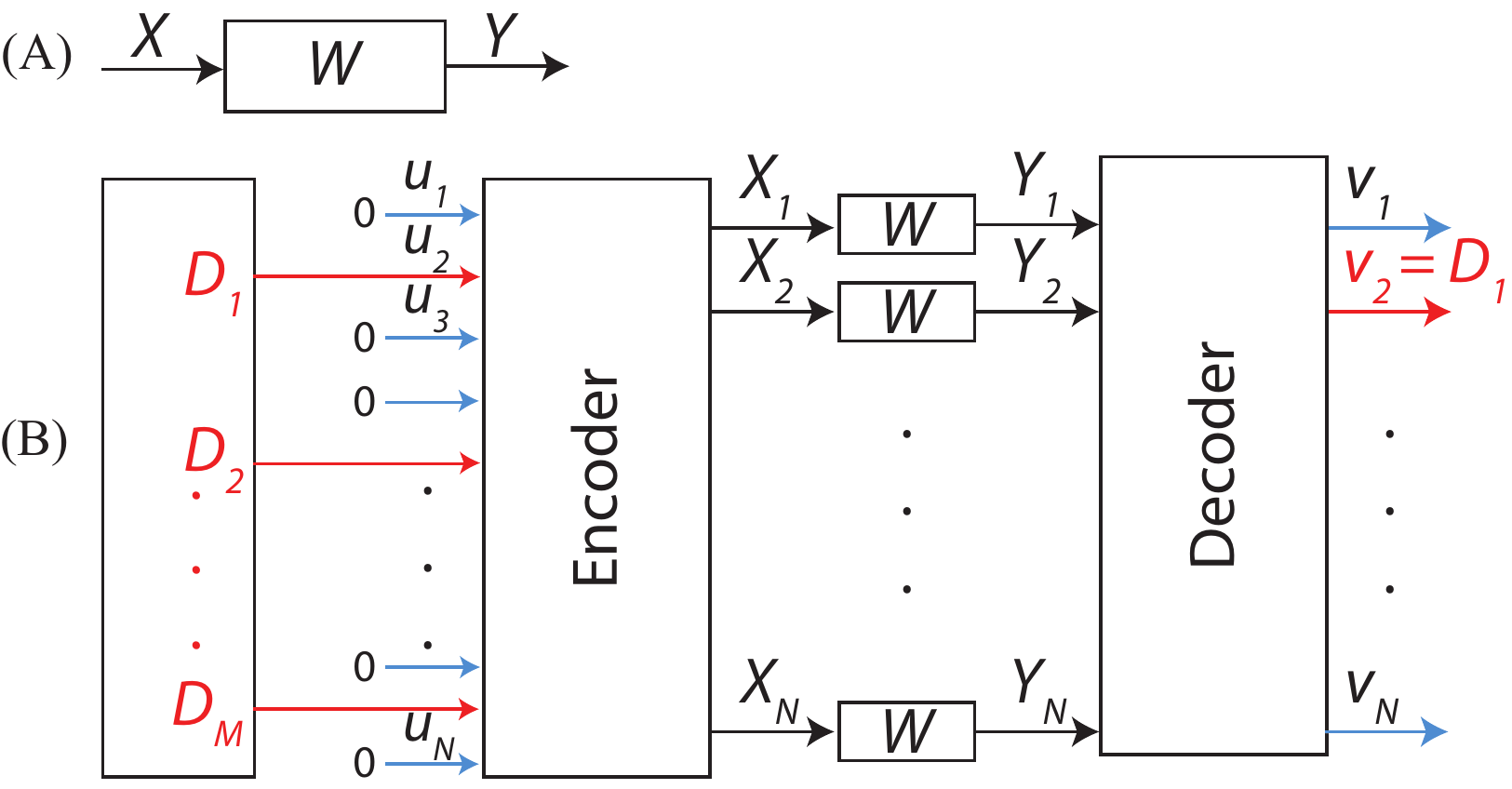}
\caption{{\bf Polar codes.} (A) A binary channel $W$ maps $X$ to $Y$. The channel has a capacity $I(W)$. (B) Polar coding comprises a preprocessing stage (Encoder) which combines data bits $D_1, D_2, \dots, D_M$ with frozen bits (say, set to 0) to create the code word $X_1X_2\dots X_N$. The transmitted code word is received as $Y_1Y_2\dots Y_N$. The “good” channels of the decoder are identified {\it a priori} and the data bits $D_1, D_2, \dots, D_M$ are read out by the receiver.}
\end{figure}

Polar codes are channel-specific codes, where the selection of good channels is a critical step for polar coding. 
The polar codes are designed by estimating and ranking the reliability of the polarized channels. 
This involves the optimization of the encoder and the decoder. 
The encoder combines the data bits with frozen bits (say, set to 0) and prepares a code word. 
An $(N, M, F)$ polar code can transmit $M$ data bits as an $N$-bit long codeword $X = X_1X_2\dots X_N$ with $|F|$ frozen bits using the following mapping:
\begin{eqnarray*}
\mathbf{X} &=& \mathbf{u G}_N,\\
\mathbf{X} &=& \mathbf{u}_{F^c} \mathbf{G}_N (F^C) + \mathbf{u}_F \mathbf{G}_N (F),
\end{eqnarray*}
where $F$ are the set of indices of the frozen bits, $F^C = \{1, 2, \dots, N\}\backslash F$ are the information bit indices, $\mathbf{u} = \{u_1, u_2, \dots, u_N\}$ are the uncoded bits and $\mathbf{G}N$ is the encoding matrix. 
The notation $\mathbf{G}_N(F)$ denotes the submatrix of $\mathbf{G}_N$ formed by the rows with indices in $F$. The encoding matrix $\mathbf{G}_N$ is defined as follows:
\begin{eqnarray*}
\mathbf{G}_N &=& \mathbf{G}_2 ^{\otimes N},\\
\mathbf{G}_2 &=& \begin{bmatrix}1 & 0 \\ 1 & 1\end{bmatrix},
\end{eqnarray*}
where $A^{\otimes N}$ is the $N$ times Kronecker product of $A$. 
The encoded bits are received as the noisy codeword $Y = Y_1Y_2\dots Y_N$, which is decoded using the successive cancellation list (SCL) decoding as $v_1v_2\dots v_N$.\cite{tal2015} 
The $M$ data bits $D_1, D_2 \dots D_M$ can be read out from the decoded bits at the good channels.

\begin{figure}
\centering
\includegraphics[width=\textwidth]{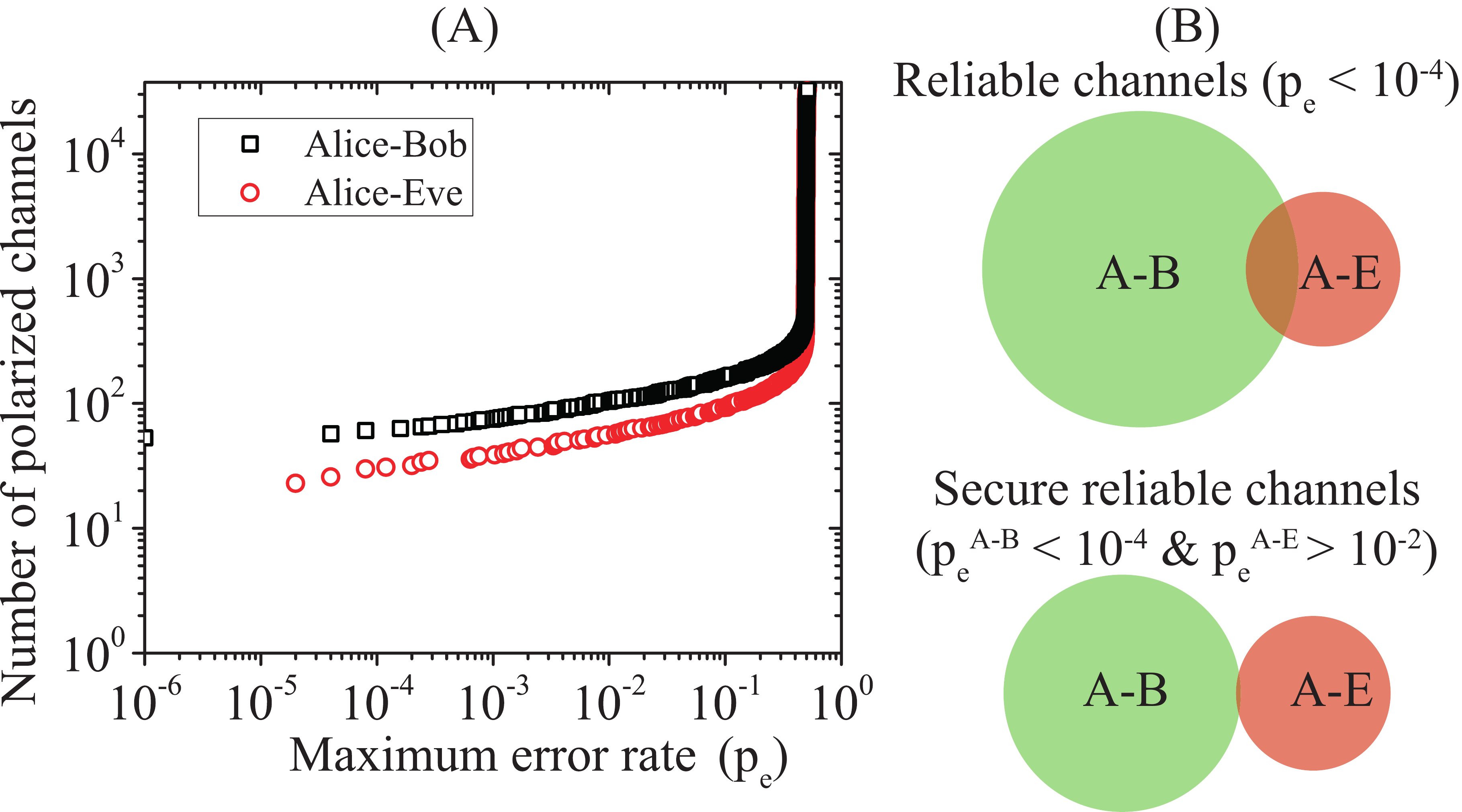}
\caption{{\bf Reliability and security of the polarized channels.} (A) shows the number of polarized channels with an error rate $p_e$ for $N = 2^{15} = 32768$ virtual channels. The error rates for the symmetric lossy channel between Alice and Bob with $alpha = 0.43$ and $\epsilon = 0.59$ and the binary symmetric channel between Alice-Eve with $\alpha = 1 - P_{\textrm{Eve}} = 0.47$ and $\epsilon = 0$ are plotted as black squares and red circles, respectively. (B) illustrates the partial overlap of reliable polarized channels (i.e. $p_e < 10^{-4})$ between Alice-Bob and between Alice-Eve. The area of the shaded regions is proportional to the channel capacities. The overlap can be reduced by selecting a subset of secure and reliable polarized channels for secret message transmission between Alice and Bob.}
\end{figure}

In our implementation, we design polar codes for $N = 2^{15}$ with a code rate $R \equiv M/N$ chosen to maximize the secret message rate between Alice and Bob, while ensuring maximal error for an eavesdropper. 
To this end, we construct the encoder and decoder for the channel between Alice and Bob with the channel parameters $\alpha = 0.43$ and $\epsilon = 0.59$. 
The reliability of the $N$ polarized channels is quantified using the error rate $p_e$ of each polarized channel, which is estimated through simulations. 
The simulations utilize density evolution methods to estimate the reliability of the Alice$\rightarrow$Bob (A-B) and the Alice$\rightarrow$Eve (A-E) channels at the encoder.\cite{mori2009}
The information block, i.e. the set of message bits, is assumed to have 16 embedded CRC (cyclic redundancy check) bits. 
We utilize a CRC-aided Successive Cancellation-List decoder (SCL) with list size of 32.\cite{tal2015} 
The Alice$\rightarrow$Eve channel has the channel parameters $\alpha = 0.46$ and $\epsilon = 0$, which are estimated using Eq. (1) assuming Eve intercepts all the signals from Alice and measures using perfect detectors. 
Further, Eve also has knowledge about the polar code design used by Alice and Bob, in addition to Bob’s public key. 
Figure 7(A) shows the number of polarized channels with the error rate $p_e$. 
The error rate $p_e$ is an upper bound on the error rate of each polarized channel, which was estimated using a density evolution method. 
At lower error rates, the ratio of the number of Alice-Bob to Alice-Eve reliable channels is seen to increase. 

We consider the channels with $p_e < 10^{-4}$ as reliable, i.e. ‘good’. As illustrated in Fig. 7(B), error correction will also allow Eve to decode using her reliable channels, with a fraction of them common between Bob and Eve. 
This will result in a loss of secrecy in the communication. 
We overcome the loss of secrecy by imposing a constraint on the choice of the reliable channels in the design of the polar code.
The selection of secure reliable channels is made such that the Alice$\rightarrow$Bob polarized channels have an error rate $p_e ^{AB} < 10^{-4}$ with the constraint that the Alice$\rightarrow$Eve channel error rate $p_e ^{AE} > 10^{-2}$. 
This results in a reduction of the number of reliable channels for transmitting message bits, i.e. a lower code rate $R$, but with minimal overlap of reliable channels between Bob and Eve. 

\begin{figure}
\centering
\includegraphics[width=\textwidth]{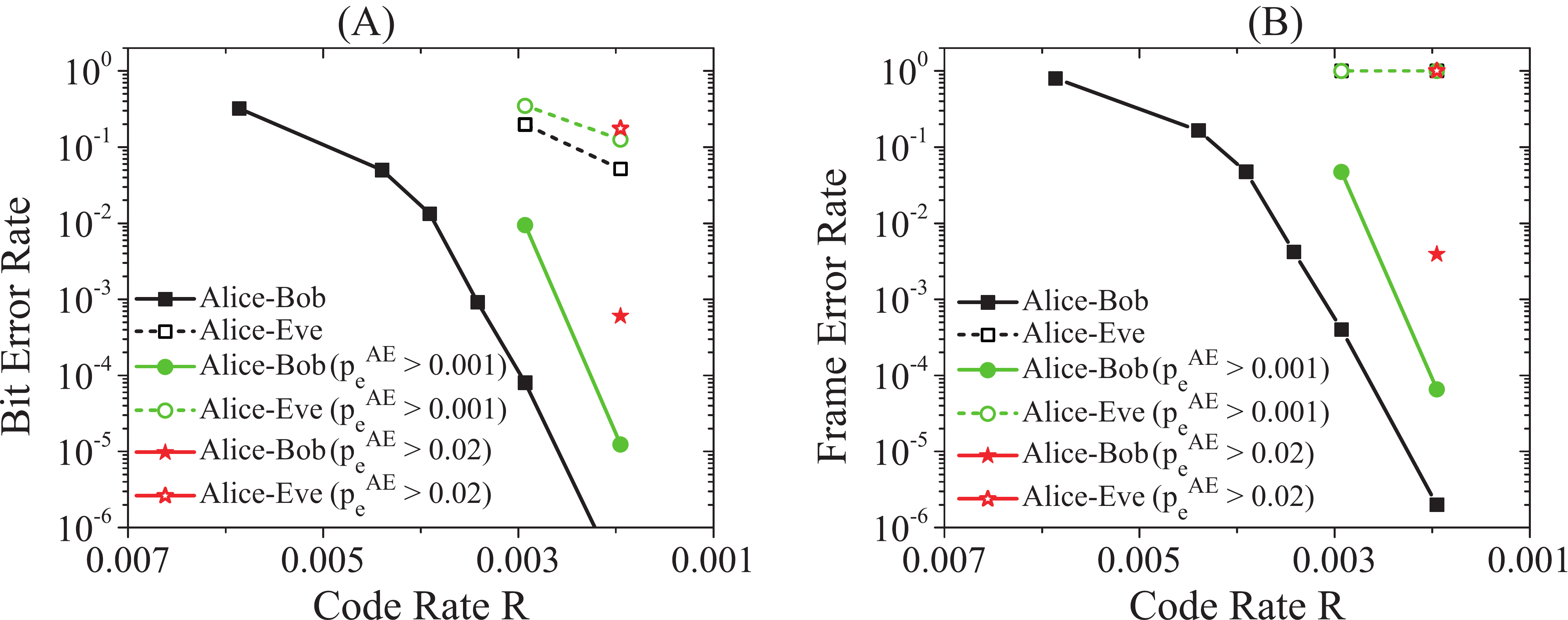}
\caption{{\bf Polar code performance.} (A) depicts the BER in the message decoded by Bob (filled symbols) and by Eve (open symbols) with varying $R$. We plot the estimated BER in communication when utilizing all the reliable channels (black squares) and when utilizing only the secure and reliable channels (green circles and red star) for different constraints on the Alice$\rightarrow$Eve error rate. (B) depicts the frame error rate FER with varying $R$ for the corresponding polar codes as shown in (A). Note that all the three curves for the Alice$\rightarrow$Eve channel are on top of each other at FER = 1.}
\end{figure}

The performance of the designed polar codes is shown in Fig. 8(A) as the bit error rate (BER) of the data bits with varying $R$. 
The BER is simulated by using a CRC-aided SCL decoding algorithm. A BER $< 10^{-4}$ for the Alice$\rightarrow$Bob channel is achieved for $R < 0.003$ with an unconstrained polar code design. 
We estimate the BER for the Alice$\rightarrow$Eve channel to be $> 0.05$ for this design. 
By imposing the reliability constraint in the selection of polarized channels, the polar codes achieve a BER for the Alice$\rightarrow$Eve channel of $> 0.12$ and for the Alice$\rightarrow$Bob channel of $10^{-5}$ at a lowered code rate $R \approx 0.002$. 
Imposing an even stricter constraint in the selection of secure reliable channels to ensure that the BER of Alice$\rightarrow$Eve channel is 0.50 will further lower the code rate, which can be prohibitive for communication. 
To understand the disparity in the information shared between Alice and Bob and Alice and Eve in a codeword transmission scheme as used here, it is necessary to consider the frame error rate FER, i.e. probability of having at least one bit error in the decoded message. 
The FER for the designed polar codes is shown in Fig. 8(B). At $R \approx 0.002$, the FER of Alice$\rightarrow$Bob channel is $< 10^{-4}$ for both the constrained $(p_e ^{AE} > 0.001)$ and the unconstrained polar code. 
The FER for the Alice$\rightarrow$Eve channel is 1, which demonstrates that for the selected code rate $R \approx 0.002$, the polar codes efficiently correct errors only on the Alice$\rightarrow$Bob channel.

\acknowledgments
 We thank Klaus Boller, Ad Lagendijk, Mehul Malik, and Willem Vos for discussions and acknowledge funding from ERC (Grant No. ERC 279248) and by a Vici grant from the Netherlands Organisation for Scientific Research (NWO). 

\bibliography{references}

\end{document}